\input{aipcheck}
\documentclass[final]{aipproc}
\layoutstyle{6x9}
\usepackage{amssymb}

\newcommand{\be}{\begin{equation}}
\newcommand{\ee}{\end{equation}}
\newcommand{\bea}{\begin{eqnarray}}
\newcommand{\eea}{\end{eqnarray}}
\newcommand{\nn}{\nonumber}
\newcommand{\slsh}[1]{{\not \! #1} }

\begin{document}

\title{The Origin of Mass~\footnote{Lectures given at the XIII Mexican School of Particles and Fields, 2-11 October, 2008, Sonora, Mexico.}}


\author{Alfredo Raya}{
  address={Universidad Michoacana de San Nicol\'as de Hidalgo, Instituto de F\'{\i}sica y Matem\'aticas. Apartado Postal 2-82, C\'odigo Postal 58040, Morelia, Michoac\'an, M\'exico}
}

\begin{abstract}
 Dynamical chiral symmetry breaking and confinement are two crucial features of Quantum Chromodynamics responsible for the nature of the hadron spectrum. These phenomena, presumably coincidental, can account for 98\% of the mass of our visible universe. In this set of lectures, I shall present an introductory review of them in the light of the Schwinger-Dyson equations.
 
\end{abstract}

\maketitle


\section{Motivation}

For millennia, mankind has always tried to explain how the world works. Every piece of knowledge, every new observation, every brilliant idea have modeled a masterpiece of that spot we like to call our home --the Universe. A common denominator of ancient cosmogonies has been to provide the elements necessary to explain what are the components of the Universe and how these ingredients sustain the delicate dance of the cosmos.  Our modern understanding about the composition of the universe reveals that most of our home is dark, either in the form of dark energy, about a 74\% of it, or as dark matter, another 22\%~\cite{DM}. Only a small portion of the Universe is what we see, and it is what has delighted us since ancient times.

Science, and particularly, physics, usually  take a reductionist path to unveil the deepest secrets of the Universe. In order to understand galaxies, we need to know everything about stars and planets: what are they made of, how they evolve in time, how  they collapse. This, in turn, requires us to learn about molecules, atoms, nuclei, nucleons, and, ultimately, quarks. Our hope is that if we manage to explain the fundamental building blocks, that is, if we manage to know what are their basic properties, we should be able to explain what is the Universe made of.  And we also need to know how these fundamental blocks are glued together as to keep the Universe flowing. These are the basic questions that elementary particle physics tries to answer.

Nature has given mankind the ability to ask questions. Most of the progress we have achieved was done answering questions,  not necessarily wise man questions; the simpler, the better. One of such has to do with one property of matter very familiar to us in our everyday experience:  \emph{Is mass a fundamental property of matter?} At first sight, that questions is rather odd: We are so used to relate mass to matter, that we cannot think of one without the other. And it is true! Our everyday experience is ruled by the classical laws of physics, elegantly summarized by Sir Isaac Newton~\cite{Newton}. In particular, Newton's second law establishes that a body accelerates proportionally to the force acting on it divided by its mass, in such a way that the motion of a body is intimately related to its mass. Furthermore, from Lavoisier we know that in every process, the mass is conserved. Thus, the Newtonian vision is that the mass of a body is the quantity of matter arising from its density and bulk conjointly. This brings us to the conclusion that the mass is, indeed, a fundamental property of matter at the classical level.

It took one more genius to ask a simple question that crumbled the Newtonian vision of mass: ``Is the energy of a body a measure of its energy content?'' Answering this question, Albert Einstein~\cite{Einstein} introduced perhaps the best known law of physics,
\be
m=\frac{E_0}{c^2}\;,
\ee
where $E_0$ is the body's rest energy. To be fair, that is not precisely the most famous form of the equation. That would be $E_0=mc^2$, which somehow prejudices to think of energy as a mass, and not the other way around. The realization of Einstein's conception of mass takes place everyday in particle accelerators, where from the collisions of particles at sufficiently high energies, much heavier particles are produced. 

The difference between the Newtonian and the Einstenian conception of mass at the fundamental level is attainable only at the subatomic level (see, for instance, Ref.~\cite{Quigg}). For example, consider the energy of a hydrogen atom in its ground state and think of it as being just its ``Einstenian'' mass. On the other hand, consider the mass of the proton and of the electron and add them in the ``Newtonian'' way. The difference of these two masses  corresponds to the ionization energy of the hydrogen, which is a difference of ${\cal O}(10^{-8})$! This tells us that at the atomic level, still the mass is Newtonian. A similar reasoning applies to the nucleus. Our empirical knowledge shows that most of the mass of a nucleus comes from the sum of the mass of its nucleons. It is hardly a $1\%$ difference, for example, the difference between the mass of a nucleus that undergoes an $\alpha$-particle emission as compared to the mass of the products of the decay. Such a difference is due to the nuclear binding energy. The mass of a nucleon, however, is a completely different story. Light quarks acquire,  through the Higgs mechanism, current masses of, say, 3-5 MeV, such that the sum of the masses of three light quarks is around 10 MeV, 98\% less than the mass of the proton! This means that the mass of a nucleon is ``Einstenian''. 

A warning note is necessary here. It is usually quoted that the Higgs mechanism~\cite{Higgs}
 (based upon the ideas of Spontaneous Symmetry Breaking of Nambu~\cite{Nambu}, whom was recently Laureated with the Nobel Prize of Physics 2008), is the origin of the mass in the Universe. This is far from being true. Although the Higgs mechanism accommodates the mass of the gauge bosons of the electroweak Standard Model respecting its gauge symmetry, and it also gives, say, electroweak masses to quarks and leptons, it certainly cannot explain around 98\% of the mass of our visible Universe, which is mostly composed of protons and nucleons. Understanding the Higgs mechanism is necesary to explain the existence of atoms and how do they form stable structures, a question as old as human kind itself. One of the physics goals of LHC~\cite{LHC} is precisely to find the Higgs boson and determine its propeties. Yet, the origin of mass of the visible Universe has an entirely different explanation.

The origin of most of the mass we can see is explained by the strong interactions of QCD~\cite{Wilczek}. The story goes as follows: Our inability to observe colored excitations in a detector tells us that valence quarks are confined inside hadrons. Furthermore,  when they are very close to each other, that is to say, when they are very energetic, they behave as free particles. This is the well known property of asymptotic freedom of QCD~\cite{Asymp}. But when they are less energetic, namely, at a distance scale of the order of a nucleon size, their respective color clouds start to overlap. In these clouds, because of the strong interactions, virtual quarks and antiquarks start condensing, forming the chiral quark condensate. This condensate provides the valence quarks of the nucleon of a dense, sticky medium in which they propagate as though their masses were some 300 MeV. Adding the mass of three of these guys explains the value of the mass of a nucleon and hence of the visible Universe! The dynamically generated mass actually measures the confining energy of the quarks inside a nucleon.  The phenomenon just described is known as the dynamical breaking of chiral symmetry, and occurs to light quarks even if their current masses vanish~\cite{Craig}. The heavier the quark, it becomes less sensitive to the dynamical effects of color interactions at low energies. Therefore, these heavy quarks acquire their masses through the Higgs mechanism instead.

In this contribution I shall present a particular \emph{Framework} to explore the details of the phenomena of confinement and dynamical chiral symmetry breaking, the Schwinger-Dyson equations~\cite{SDE}, sometimes also referred to as Dyson-Schwinger equations. For this purpose, I shall work out a toy model of QCD which, however, is interesting by own merits, QED$_3$, that is, the standard quantum electrodynamics restricted to a plane. Students are encouraged to follow the \emph{Approach} and develop the necessary techniques to unveil the secrets of the origin of mass.

\section{Framework}

A natural framework to study the strong interactions of QCD is, indeed, a lattice simulation. Lattice gauge theory has provided an accurate description of the hadron spectrum, within a 10\% error of the experimental data~\cite{CP-PACS}, which confirms the physical picture of the origin of mass described above~\cite{Wilczek}. A different, and in many senses, complementary platform to study non perturbative phenomena in Quantum Field Theories is provided by the Schwinger-Dyson equations (SDEs)~\cite{SDE}. These are the field equations of a given theory, and form an infinite tower of relations among the Green's functions of the theory. In order to address the issues of confinement and dynamical chiral symmetry breaking (DCSB), and for the sake of simplicity, instead of solving the SDEs for QCD, I shall restrict myself to a simpler and more familiar theory, QED, starting from the Lagrangian
\be
{\cal L}=\bar\psi (i\gamma^\mu (\partial_\mu-ieA_\mu)-m_0)\psi -\frac{1}{4}F_{\mu\nu} F^{\mu\nu} - \frac{1}{2\xi}(\partial_\mu A^\mu)^2\;,
\ee
where all the quantities carry their usual meaning.
\begin{center}
\begin{figure}[t]
\includegraphics[width=0.8\textwidth]{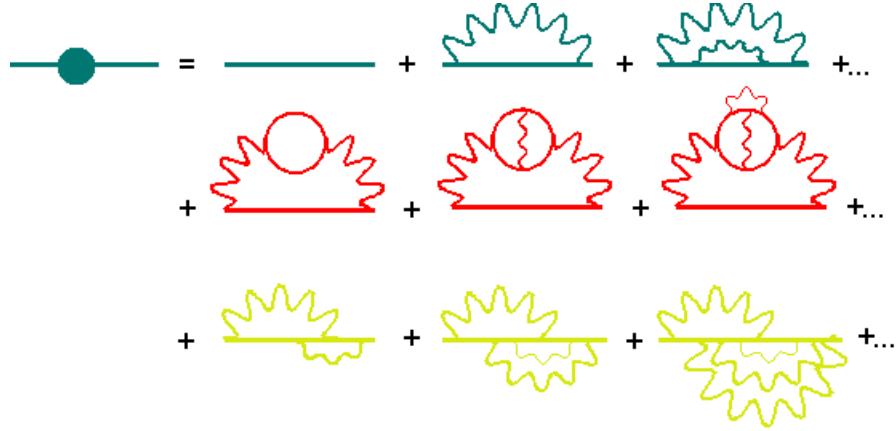}
\caption{Perturbative Expansion of the Fermion Propagator}
\label{fig:PT}
\end{figure}
\end{center}

\begin{center}
\begin{figure}[t]
\includegraphics[width=0.3\textwidth]{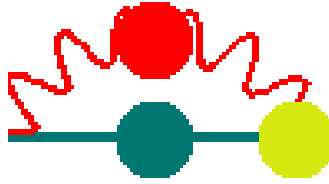}
\caption{Fermion Self Energy $\Sigma(p)$}
\label{fig:sigma}
\end{figure}
\end{center}

\begin{center}
\begin{figure}[t]
\includegraphics[width=0.8\textwidth]{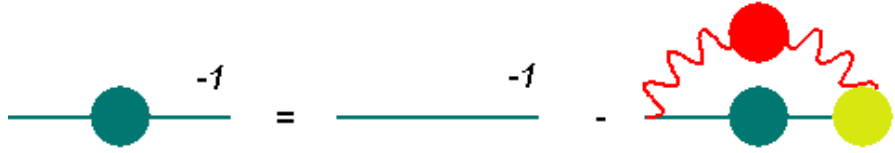}
\caption{Schwinger-Dyson equation for the Inverse Fermion Propagator in QED}
\label{fig:SDEfer}
\end{figure}
\end{center}

In order to understand the physical content of the SDEs (see, for instance~\cite{Pennington,Nova}), consider the perturbative expansion of the fermion propagator $S_F(p)$ in QED, Fig. (\ref{fig:PT}). There appear three types of radiative corrections, to the fermion propagator itself, to the photon propagator and to the fermion-photon vertex. Defining the self-energy as in Fig.~(\ref{fig:sigma}), all these corrections are incorporated in this object, in such a fashion that the perturbative expansion in terms of $\Sigma(p)$ now reads
\be
S_F(p)=S_F^{(0)}(p)+S_F^{(0)}(p)\Sigma(p)S_F^{(0)}(p)+S_F^{(0)}(p)\Sigma(p)S_F^{(0)}(p)\Sigma(p)S_F^{(0)}(p)+\ldots
\ee
\begin{center}
\begin{figure}[t]
\includegraphics[width=0.8\textwidth]{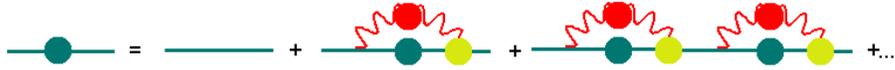}
\caption{Perturbative Expansion of the Fermion Propagator in terms of $\Sigma(p)$}
\label{fig:PTsigma}
\end{figure}
\end{center}
and is shown in Fig.(\ref{fig:PTsigma}). Here, $S_F^{(0)}(p)$ is the bare fermion propagator. Factorizing $S_F^{(0)}(p)$, the remaining expression is a geometric series in $\Sigma(p) S_F^{(0)}(p)$, which adds to
\be
S_F(p)=\frac{S_F^{(0)}(p)}{1-\Sigma(p)S_F^{(0)}(p)}\;.\label{SD1}
\ee
Equivalently, leaving the bare propagator apart and factorizing $S_F^{(0)}(p)\Sigma(p)S_F^{(0)}(p)$, we again reach at a geometric series in $\Sigma(p)S_F^{(0)}(p)$, and hence
\be
S_F(p)=S_F^{(0)}(p)+S_F^{(0)}(p)\Sigma(p)\frac{S_F^{(0)}(p)}{1-\Sigma(p)S_F^{(0)}(p)}\;.\label{SD2}
\ee
Comparing eqs.~(\ref{SD1}) and~(\ref{SD2}), we obtain
\be
S_F(p)=S_F^{(0)}(p)+S_F^{(0)}(p)\Sigma(p)S_F(p)\;.
\ee
This is the SDE for the fermion propagator. On multiplying this expression by $S_F^{(0)\ -1}(p)$ from the left and $S_F^{-1}(p)$ from the right, we arrive at the most familiar SDE for the inverse fermion propagator
\be
S_F^{-1}(p)= S_F^{(0)\ -1}(p) -\Sigma(p)\;,
\ee
and is depicted in Fig.~(\ref{fig:SDEfer}). 
This corresponds to the expression
\be
S_F^{-1}(p)=S_F^{(0)\ -1}(p) -ie^2\int \frac{d^4k}{(2\pi)^4} \gamma^\mu S_F(k) \Gamma^\nu(k,p) \Delta_{\mu\nu}(k-p)\;,\label{SDE1}
\ee
where $e^2$ is the electromagnetic coupling, $\Delta_{\mu\nu}$ represents the complete photon propagator and $\Gamma^\nu$ the full fermion-photon vertex. This is the equation we need to solve in order to study DCSB.

\begin{center}
\begin{figure}[t]
\includegraphics[width=0.8\textwidth]{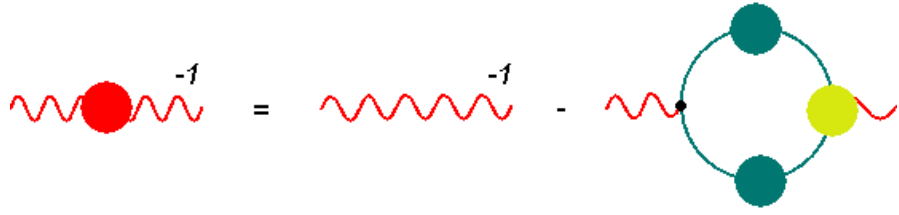}
\caption{Schwinger-Dyson equation for the Photon Propagator  in QED}
\label{fig:SDEphot}
\end{figure}
\end{center}

\begin{center}
\begin{figure}[t]
\includegraphics[width=0.8\textwidth]{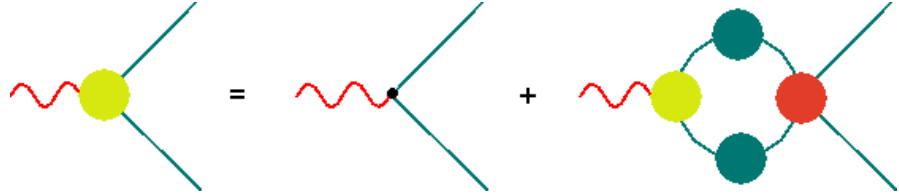}
\caption{Schwinger-Dyson equation for the Fermion-Boson Vertex in QED}
\label{fig:SDEfbv}
\end{figure}
\end{center}
Notice that the Green's functions involved in $\Sigma(p)$ obey their own SDE. The SDEs for the photon propagator and the fermion-photon vertex are depicted in Figs.~(\ref{fig:SDEphot}) and~(\ref{fig:SDEfbv}), respectively. Observe that the two-point Green's functions are coupled to one another and to a three-point function, which, in turn, is coupled to them and to a four-point function and so on. Hence, the set of SDEs, which are the field equations of QED, form an infinite tower of relations among the Green's functions of the theory.

Now, we need to find out which kind of unknowns we are dealing with before we even try to find them. Let us notice that  the complete fermion propagator, $S_F(p)$, can be expressed in terms of two unknown scalar functions of the momentum squared, multiplied each by the tensor structures ${\mathbf 1}$ and $\slsh{p}$ which appear in the Dirac equation. For reasons that will become clear shortly, we  write the propagator as
\be
S_F(p)=\frac{F(p)}{\slsh{p}-M(p)}\;,
\ee
and refer to $F(p)$ as the fermion wavefunction renormalization and $M(p)$ as the mass function. Notice that the pole of the propagator, i.e., the mass of the particle, is located at $p^2=M^2(p)$. The bare propagator $S_F^{(0)}(p)$ corresponds to $F(p)=1$ and $M(p)=m_0$. The photon propagator, in turn, can be written as
\be
\Delta^{\mu\nu}(q)=\frac{G(q)}{q^2}\left(g^{\mu\nu}-\frac{q^\mu q^\nu}{q^2} \right) +\xi \frac{q^\mu q^\nu}{q^4}\;,
\ee
where $G(q)$ represents the unknown photon wavefunction renormalization and $\xi$ is the usual covariant gauge parameter. $\xi=0$ labels the Landau gauge, and $\xi=1$ the Feynman gauge. The bare propagator $\Delta_{\mu\nu}^{(0)}(q)$ is obtained setting $G(q)=1$. Finally, the fermion-photon vertex can be expressed as
\be
\Gamma^\mu(k,p)=\sum_{i=1}^{12} v_i(k,p) V_i^\mu\;,
\ee
where $v_i(k,p)$ are unknown scalar functions and $V_i^\mu$ are the basis vectors formed from the products of each the three vectors $k^\mu$, $p^\mu$ and $\gamma^\mu$ with each of the 4 spin structures ${\mathbf 1}$, $\slsh{k}$, $\slsh{p}$ and $\slsh{k} \slsh{p}$. In summary, there are 15 unknown functions in eq.~(\ref{SDE1}). Fortunately, the gauge principle tells us that not all of them are independent, as they are related through gauge identities like Ward-Green Takahashi identities~\cite{WGTI} and so on. A recent review on different truncations of SDE can be found in~\cite{Nova}.

Confinement and DCSB are usually studied from the fermion propagator. This means that  if we want to truncate the tower of SDEs at the level of eq.~(\ref{SDE1}), reasonable assumptions have to be made on the form of the photon propagator and the fermion-photon vertex. Below we shall consider one of such truncation schemes which, on one hand, makes the analysis very neat and, on the other hand, shows the key features of QCD that explain the origin of mass.

\section{Approach}

In order to study DCSB and confinement, let us work in a toy model of QCD: QED$_3$. That is, let us work with the ordinary QED, but restricted to a plane. Such a theory is super-renormalizable, so that we don't have to deal with regularization issues in a truncation of SDEs (see, for example, Ref.~\cite{Nova} for the implications of this fact). Contrary to the ordinary QED, the planar theory exhibits confinement~\cite{ConfQED3}, in a manner such that it resembles more to QCD that to QED itself. Even more, it has been argued very recently that both the restoration of chiral symmetry and the confinement/deconfinement phase transitions take place simultaneously in this theory~\cite{simult}. Furthermore, to the corresponding Lagrangian we can add a Chern-Simons term, which induces a topological, gauge invariant mass to the photons, making the structure of the theory even richer (a discussion of the most general QED$_3$ Lagrangian can be found in Ref.~\cite{zero}). QED$_3$ has multiple applications not only because it emerges as the infinite temperature limit of QED~\cite{HTqed}, but there are many condense matter systems, ranging from high-T$_c$ superconductors~\cite{supercon} the quantum Hall effect~\cite{zero,QHE}, and more recently, graphene~\cite{graphene}, described by this theory. In the field of DCSB, QED$_3$ has provided a popular battleground for lattice, SDEs  and other studies, particularly  regarding a possible critical number of fermion families for which the phenomenon ceases to take place~\cite{DCSB}.

\begin{center}
\begin{figure}[t]
\includegraphics[width=0.8\textwidth]{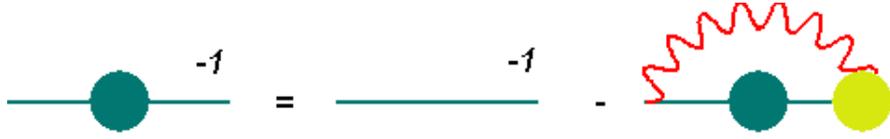}
\caption{Quenched truncation of the Schwinger-Dyson equation for the Fermion Propagator in QED}
\label{fig:SDEquenched}
\end{figure}
\end{center}

We start our study of DCSB from the SDE for the quark propagator $S_F(p)$, which in QED$_3$ is given by
\be
S_F^{-1}(p)=S_F^{(0)\ -1}(p) -ie^2\int \frac{d^3k}{(2\pi)^3} \gamma^\mu S_F(k) \Gamma^\nu(k,p) \Delta_{\mu\nu}(k-p)\;.
\ee
Here, $e^2$ is the dimensionful coupling of the theory that sets its natural scale. This means that any new mass scale, like the dynamical mass, should be proportional to $e^2$.  To emphasize the phenomenon of DCSB, we consider massless fermions to start with, setting $m_0=0$ in the bare propagator. In order to truncate the tower of SDEs at this level, we need ans\"atze for $\Gamma^\nu$ and $\Delta_{\mu\nu}$. Let us start by in neglecting fermion loops, $\Delta_{\mu\nu}\to \Delta_{\mu\nu}^{(0)}$. This is the quenched approximation of QED, which is pictorially represented in Fig.~(\ref{fig:SDEquenched}). 

\begin{center}
\begin{figure}[t]
\includegraphics[width=0.8\textwidth]{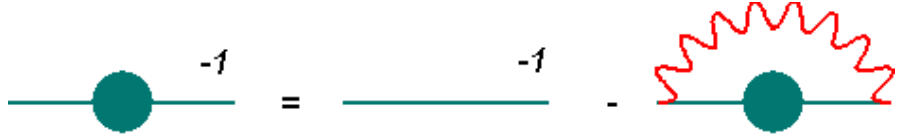}
\caption{Rainbow truncation of the Schwinger-Dyson equation for the Fermion Propagator in QED}
\label{fig:SDEeainbow}
\end{figure}
\end{center}

With a suitable choice of the fermion-photon vertex, the SDE for the fermion propagator can be solved self-consistently. Possible the simplest interaction we can consider is the bare interaction, $\Gamma^\nu(k,p)=\gamma^\nu$. This is the so-called rainbow approximation, depicted in Fig.~(\ref{fig:SDEeainbow}), and  under which the gap equation reads
\be
\frac{{\slsh{p}}-M(p)}{F(p)} = \slsh{p} -\frac{i\alpha}{2\pi^2} \int d^3k \frac{F(k)}{k^2-M^2(k)} \gamma^\mu (\slsh{k}+M(k)) \gamma^\nu \Delta_{\mu\nu}^{(0)}(k-p)\;.\label{rainbow}
\ee
Here we have set $\alpha=e^2/(4\pi)$, as usual. This is a matrix equation which can be converted into a system of scalar integral equations for $F(p)$ and $M(p)$ after multiplying it by $\slsh{p}$ and ${\mathbf 1}$. However, in the Landau gauge  $F(p)=1$ (see, for example, Ref.~\cite{rainbow}), so, in this gauge we are left with a single equation for $M(p)$ alone.  Such an equation is derived taking the trace to eq.~(\ref{rainbow}) with $F(p)=1$, yielding
\be
M(p)=\frac{i\alpha}{2\pi^2} \int d^3k \frac{M(k)}{k^2-M^2(k)} g^{\mu\nu} \Delta_{\mu\nu}^{(0)}(k-p)\;.
\ee
After contracting with the photon propagator, we carry out a Wick rotation to Euclidean space with the following prescriptions,
\bea
(k^0, \ p^0) &\to& (ik^0_E, \ ip^0_E)\;,\nn\\
(k^2,\ p^2,\ (k-p)^2)&\to & (-k^2_E,\ -p^2,\ -(k-p)^2_E)\;,\nn\\
\int d^3k &\to& i\int d^3k_E\;.
\eea
For ease of notation, we shall drop the subscript $E$ from every Euclidean space expression. Hence, we are finally left with the following expression for the Euclidean gap equation in QED$_3$,
\be
M(p)=\frac{\alpha}{\pi^2} \int \frac{d^3k}{(k-p)^2} \frac{M(k)}{k^2+M^2(k)}\;.
\ee
We can still go further making reasonable simplifying assumptions~\cite{Maris}. Since the DCSB is an infrared phenomenon, we expect the dynamically generated mass to serve as an IR cut-off for the gap equation. We can then linearize the above expression setting in the denominator of the integrand
\be
M^2(k)=M^2(0)\equiv m_{dyn}^2\;,
\ee
which yields the linearized gap equation
\be
M(p)=\frac{\alpha}{\pi^2} \int \frac{d^3k}{(k-p)^2} \frac{M(k)}{k^2+m_{dyn}^2}\;, \label{SDEq}
\ee
Notice that $M(p)=0$ is a (trivial) solution of this equation, and would correspond to that derived in perturbation theory. However, we are interested in a non trivial solution, which we shall look for making use of both analytical and numerical techniques.

\subsection{Analytical treatment}

In order to solve eq.~(\ref{SDEq}) analytically, define the function $\chi(p)$ through~\cite{Maris}
\be
M(p)=(p^2+m_{dyn}^2)\chi(p)\;.
\ee
Then we are left with
\be
(p^2+m_{dyn}^2)\chi(p) = \frac{\alpha}{\pi^2} \int \frac{d^3k}{(k-p)^2}\chi(k)\;.
\ee
Now, defining the Fourier transform of $\chi(p)$,
\be
\varphi(r)= \int \frac{d^3p}{(2\pi)^3}\chi(p)e^{ip\cdot r}\;,
\ee
we observe that it satisfies the differential equation
\be
\frac{d^2}{dr^2}\varphi(r) + \frac{2}{r}\frac{d}{dr}\varphi(r) + \left(m^2-\frac{2\alpha}{r} \right)\varphi(r)=0\;,
\ee
whose solution is
\begin{center}
\begin{figure}[t]
\includegraphics[width=0.8\textwidth,angle=-90]{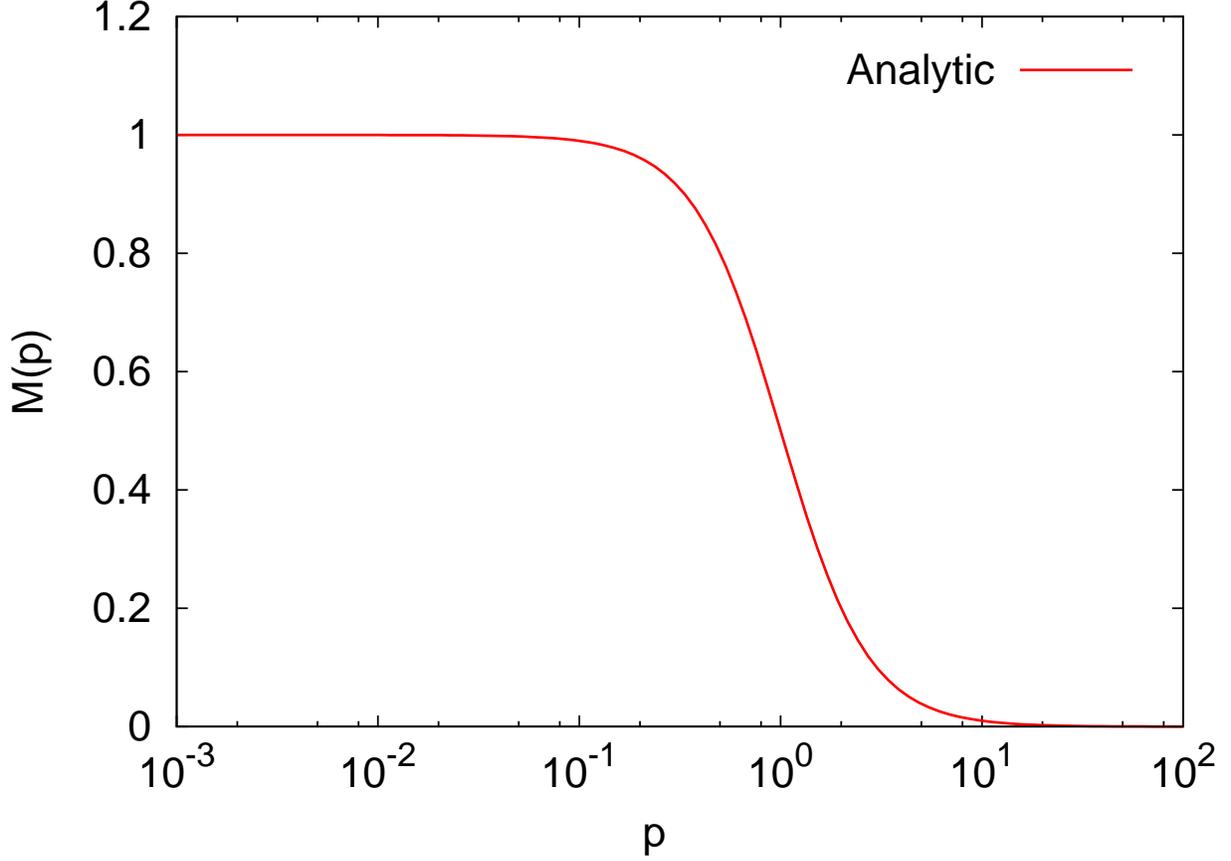}
\caption{Analytical solution to the linearized Schwinger-Dyson equation for the Fermion Propagator in QED$_3$. The scale of the graph is set by $m_{dyn}=1$}
\label{fig:analyt}
\end{figure}
\end{center}
\be
\varphi(r)=C e^{-m_{dyn} r}.
\ee
On Fourier transforming back the result to momentum space, the normalization constant $C$ is found demanding the infrared behavior of  $M(p\to 0)=m_{dyn}$. The resulting mass function is then
\be
M(p)=\frac{m_{dyn}^3}{p^2+m_{dyn}^2}\;,\label{mp}
\ee
and is depicted in Fig.~(\ref{fig:analyt}).
We observe that the mass function  falls-off as $M(p\to\infty)\sim 1/p^2$. This is consistent with the expression for the chiral condensate,
\be
\langle\bar\psi\psi\rangle= \lim_{p\to\infty} \frac{p^2 M(p)}{2}\;,
\ee
derived from the Operator Product Expansion (see, for example,~\cite{rainbow}), since it demands the ultraviolet behavior $M(p)\sim 1/p^2$ in order to render the condensate momentum independent. These limits will become relevant to test the validity of the numerical procedure that we are about to develop below.

\subsection{Numerical treatment}

Consider again eq.~(\ref{SDEq}) and select a reference frame where the angle between  momenta $k$ and $p$ is $\theta$. Then we can write $d^3k = k^2dk \sin\theta d\theta d\phi$ for $\phi\in (0,2\pi)$, $\theta \in (0,\pi)$ and $k\in(0,\infty)$. The integration over $\phi$ gives a factor of $2\pi$, and the integration over $\theta$ is 
\be
\int_0^\pi \frac{d\theta\ \sin\theta}{k^2+p^2-2kp\cos\theta}=\frac{1}{2kp}\ln{\left| \frac{k+p}{k-p}\right|}\;.
\ee
Gathering things together, we have
\be
M(p)=\frac{\alpha}{\pi p} \int_0^\infty dk \frac{k M(k)}{k^2+m_{dyn}^2}\ln{\left| \frac{k+p}{k-p}\right|}\;.
\ee
Observe that there is an integrable singularity in the logarithm for $k=p$. In order to avoid numerical complications coming from this singularity, we can approximate 
\be
\ln{\left| \frac{k+p}{k-p}\right|} \simeq \frac{2p}{k}\vartheta(k-p)+\frac{2k}{p}\vartheta(p-k)\;,
\ee
where $\vartheta(x)$ is the step function. Such an approximation is valid for $k\gg p$ and $k \ll p$, and yields the simplified SDE given by
\be
M(p)=\frac{2\alpha}{\pi p} \int_0^\infty dk \frac{k M(k)}{k^2+m_{dyn}^2}\Bigg\{ \frac{p}{k}\vartheta(k-p)+\frac{k}{p}\vartheta(p-k)\Bigg\}\;.
\ee
This is the equation we need to solve numerically. 

Set $e^2=1$ to start with. Then, select an integration interval $(\kappa,\lambda)$ such that 
\be
M(p)\simeq\frac{2\alpha}{\pi p} \int_\kappa^\lambda dk \frac{k M(k)}{k^2+m_{dyn}^2}\Bigg\{ \frac{p}{k}\vartheta(k-p)+\frac{k}{p}\vartheta(p-k)\Bigg\}\;.\label{iesde}
\ee
This integral equation is of the form
\be
M(p)= \int_\kappa^\lambda dk \ {\rm Ker}(k,M(k);p)\;,
\ee
where ${\rm Ker}(k,M(k);p)$ is the kernel of the equation. Thus, we need a quadrature rule which allows us to approximate
\be
M(p)\simeq \sum_{i=1}^{n_{max}} \Delta_i \ {\rm Ker}(k_i, M_i;p)\;,
\ee
\begin{center}
\begin{figure}[t]
\includegraphics[width=0.8\textwidth,angle=0]{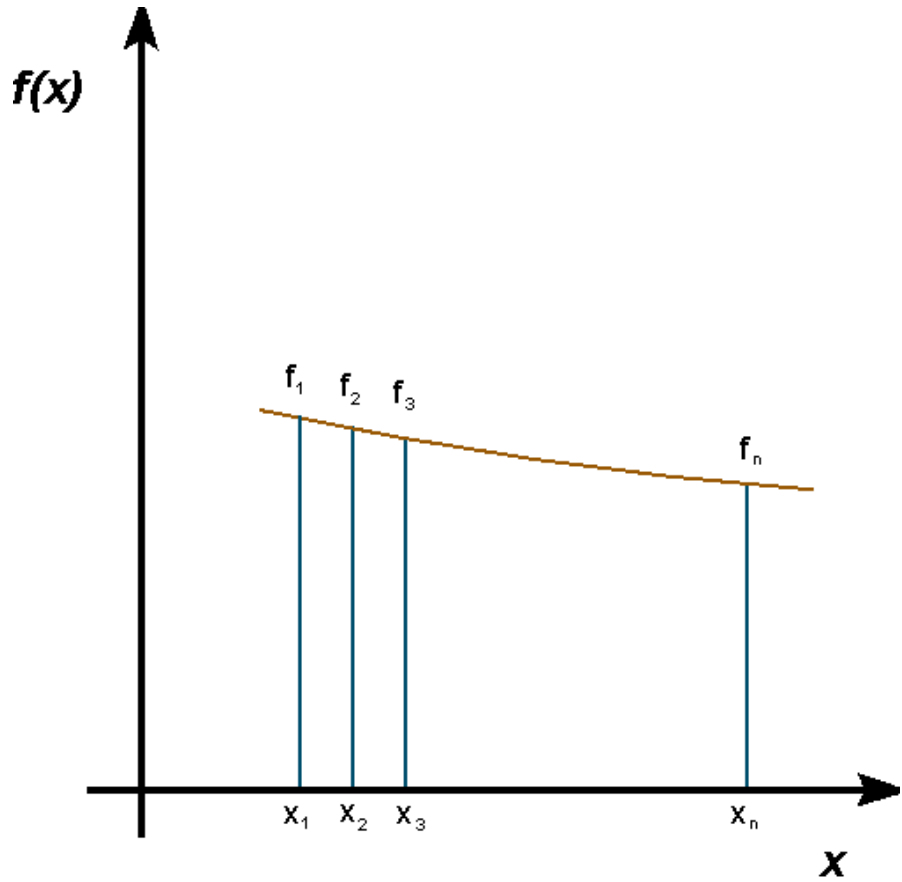}
\caption{A quadrature rule for integration}
\label{fig:quadra}
\end{figure}
\end{center}
where the $\Delta_i$ are the weights of the quadrature and $k_i$ the corresponding nodes, as shown in Fig.~(\ref{fig:quadra}). Here we have used the notation $M_i=M(k_i)$. In principle, we should be interested in the solution of the above expression in the interval $p\in (\kappa,\lambda)$.  However, we can resort to the collocation techniques to solve this equation, imagining that we are only interested in the solution $M(p)$ over a finite set of grid points $p_1,\ldots,p_j$ rather that the solution over the complete interval. In such a case, we reach at an equation like
\be
M(p_j)= \sum_{i=1}^{n_{max}} \Delta_i \ {\rm Ker}(k_i, M_i;p_j)\;,
\ee
for each $p_j$, with the added advantage that if the set of grid points $\{p_j\}$ is selected as the same set of points of the quadrature rule, i.e. $p_j=k_j$ for $j=1,\ldots,n_{max}$, we are left with a system of $n_{max}$ algebraic equations for the $n_{max}$ unknowns $M_i$. This is the feature of collocation techniques, an integral equation over an interval is translated into an algebraic system of equations.

Let us apply these techniques to solve eq.~(\ref{iesde}) in a convenient interval $(\kappa,\lambda)$. A good integration interval, will be one for which, as $\kappa\to 0$  and $\lambda \to\infty$,  both the infrared and ultraviolet behaviors of the mass function we discussed previously have settled over a couple of orders of magnitude of momentum. For concreteness, let us select the interval $(10^{-4},10^3)$. The next step is to select a quadrature rule to perform the integration. The simplest choice would be a trapezoidal rule and thus a set of grid points in the selected interval. A logarithmic grid is the most appropriate, since there are several orders of magnitude involved. Furthermore, one has to make sure that the integration points are uniformly distributed over the interval of integration. It is desirable to have the same number of grid points over each decade in the interval. For example, if we want to distribute $\mu$ points in each decade of the interval, and we want   
\be
n_{max}=[\log_{10}(\lambda)-\log_{10}(\kappa)]\mu+1\;
\ee
grid points, defining
\be
\eta= \frac{\ln{(10)}}{\mu}\;,
\ee
we observe that the recursion
\be
p_{j+1}=p_j+\kappa e^{(j-1)\eta}(e^\eta-1)\;, \qquad j=1,\ldots,n_{max}-1
\ee
provides us with a set of points uniformly distributed over the interval, exactly $\mu$ of them in each decade.

We now need to construct the integration weights of our trapezoidal quadrature. Notice that each weight $\Delta_j$ in the logarithmic grid should depend of the grid point $p_j$, as opposed to the ordinary ``equally spaced'' weights of a linear grid. Therefore, we can define them as
\bea
\Delta_j &=& \frac{1}{2}\left(p_{j+1}-p_{j-1}\right)\,\qquad j=2,\ldots,n_{max}-1\nn\\
\Delta_1 &=& \frac{1}{2}\left(p_{2}-p_{1}\right)\;,\qquad
\Delta_{n_{max}} \ =\  \frac{1}{2}\left(p_{n_{max}}-p_{n_{max}-1}\right)\;.
\eea
With these ingredients, we are ready to write down the system of equations to be solved. From the analytical treatment, we expect $M(p)$ to behave like a constant as $p\to 0$. That means that in the denominator of eq.~(\ref{iesde}), we can write $m_{dyn}=M_1$, and thus, for $j=1,\ldots,n_{max}$, we have an equation of the form
\be
M_j=\frac{2\alpha}{\pi p} \sum_{i=1}^{n_{max}} \Delta_i \ \frac{k_i M_i}{k_i^2+M_1^2}\Bigg\{ \frac{k_j}{k_i}\vartheta(i-j)+\frac{k_i}{k_j}\vartheta(j-i)\Bigg\}\;, \label{system}
\ee
since if $i>j$, then $k_i>k_j$. We can now proceed to solve the system of equations given above by standard techniques, like the Newton-Raphson method. 
\begin{center}
\begin{figure}[t]
\includegraphics[width=0.8\textwidth,angle=-90]{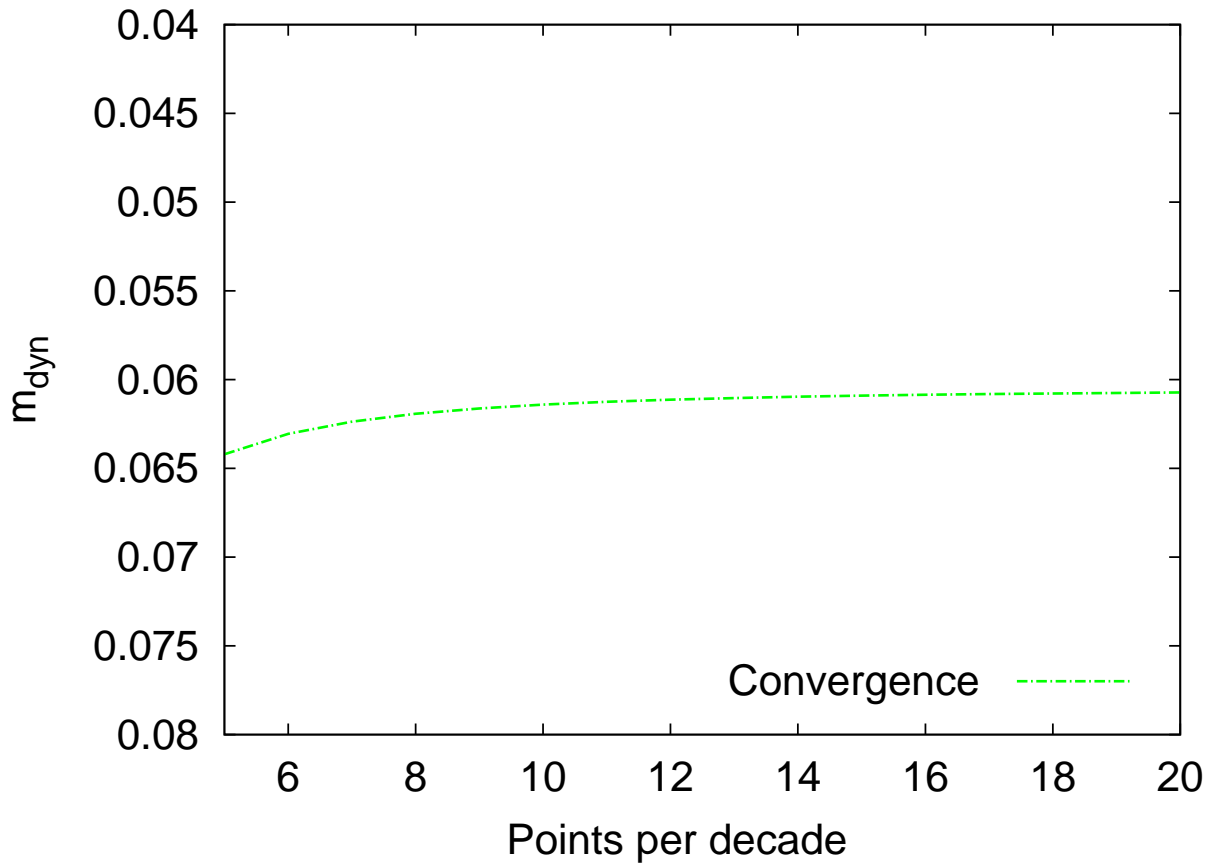}
\caption{Stability test of the numerical solution with $m_{dyn}$}
\label{fig:stab}
\end{figure}
\end{center}

\begin{center}
\begin{figure}[t]
\includegraphics[width=0.8\textwidth,angle=-90]{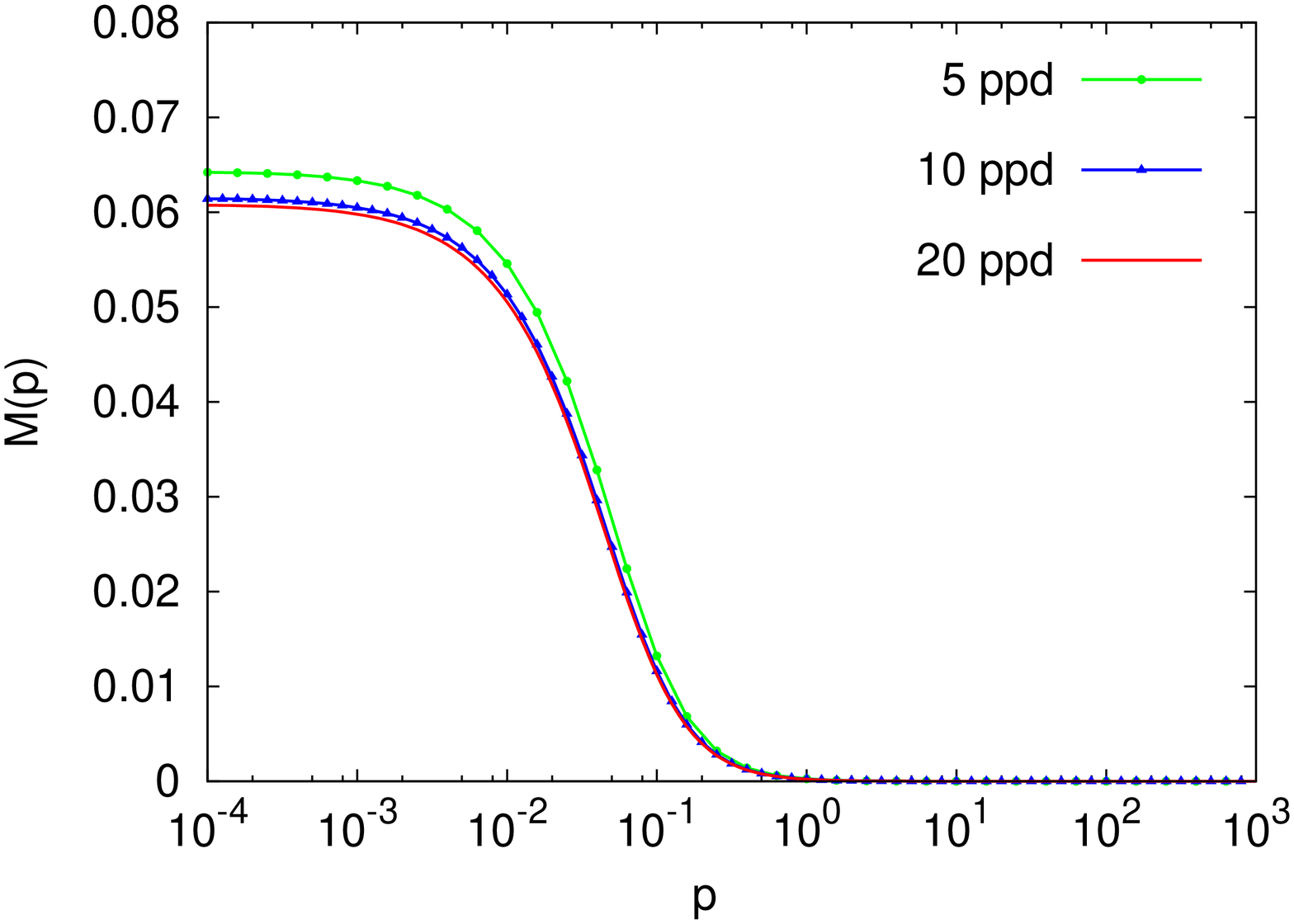}
\caption{Mass function for various points per decade}
\label{fig:mass}
\end{figure}
\end{center}

\begin{center}
\begin{figure}[t]
\includegraphics[width=0.8\textwidth,angle=-90]{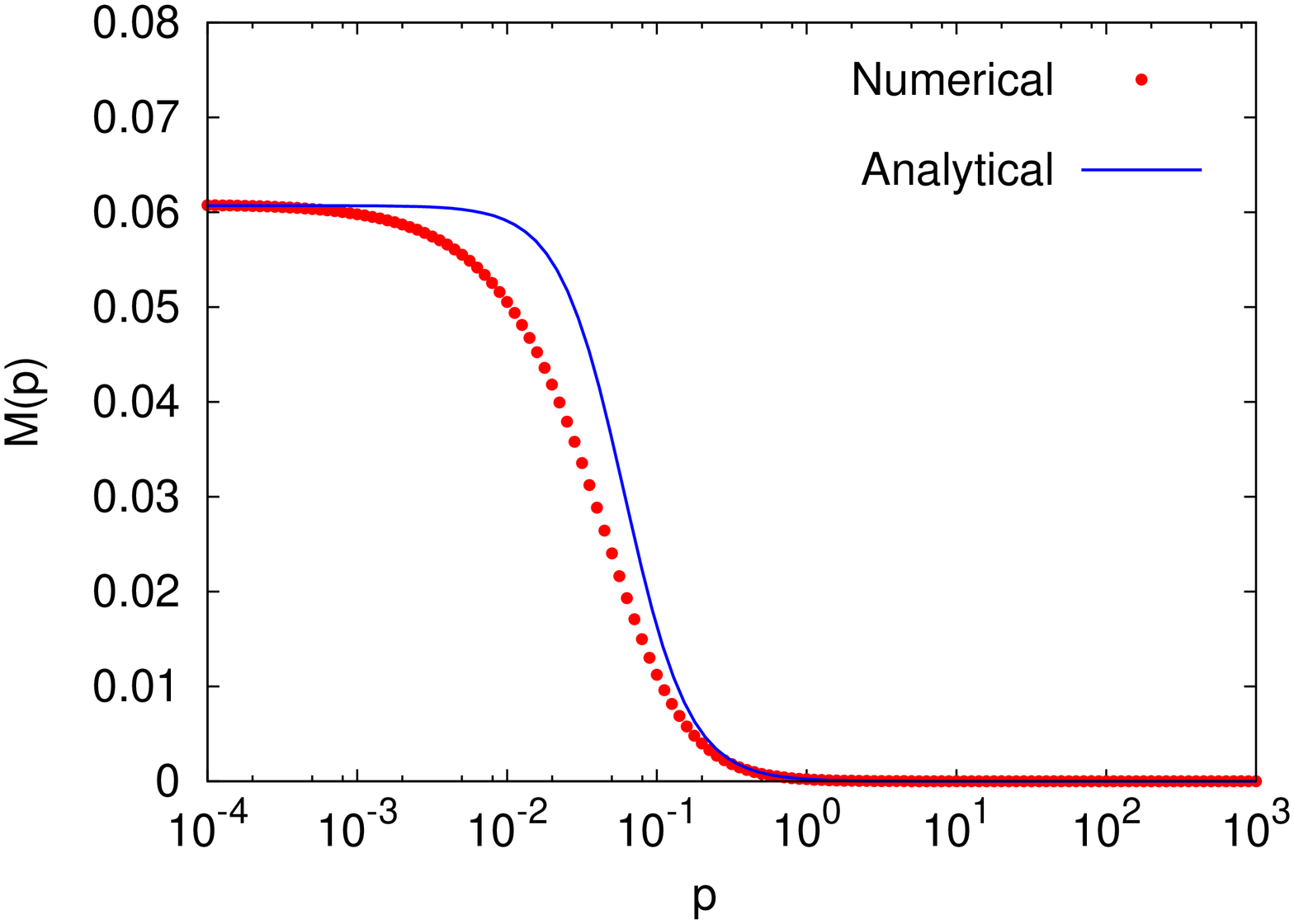}
\caption{Analytical vs. Numerical comparison of the Mass function}
\label{fig:avsn}
\end{figure}
\end{center}

In order to test the stability of the procedure, we should establish the dependence of the solution upon the number of grid points. In Fig.~(\ref{fig:stab}) we depict the behavior of $m_{dyn}$ as a function of $\mu$, and observe that for $\mu>8$, the variation of $m_{dyn}$ with the number of points per decade is negligible. In Fig.~(\ref{fig:mass}), we depict the mass function $M(p)$ for different values of $\mu$. We observe precisely that for $\mu\gtrsim 10$, all the curves lay practically on top of each other, which points out the fast stability of the numerical procedure. Figure~(\ref{fig:avsn}) shows de comparison of the analytical expression, eq.~(\ref{mp}), and the numerical result for $\mu=20$. The agreement of the curves  both in the infrared and ultraviolet allows us to readily infer that both the infrared constant behavior of $M(p)$ and the ultraviolet fall-off as $1/p^2$. The difference, of course, arises in the intermediate region, where the effects of the linearization are stronger.

Although from this exercise we conclude that QED$_3$ is a theory that supports DCSB, we still need to find out whether the solution to the SDE under the set of simplifying assumptions, eq.~(\ref{mp}), supports confinement. Such an issue is discussed below.

\subsection{Confinement Test}

Confinement is connected to our impossibility of detecting colored states. From the theoretical point of view, there are several criteria connected to the infrared behavior of Green's functions to establish confinement (see, for example,~\cite{Fischer}), like the Kugo-Ojima criterion~\cite{kugo} and the Gribov-Zwanzinger confinement scenario~\cite{gz}, which demand an infrared suppressed behavior of the gluon propagator and quark-gluon vertex and an infrared enhancement of the ghost propagator in QCD. Recent lattice simulations, however, seem to disagree with those criteria, particularly in the behavior of the ghost and gluon propagators~\cite{disa}. Here, we adopt a different point of view to establish confinement in QED$_3$, based upon the Axiom of Reflection Positivity~\cite{ARP}, which establishes that if an Euclidean Green's function describes a stable excitation, it should be positive definitive. This statement provides us with a confinement test that would help us understand this phenomenon in QED$_3$ and its connection with DCSB~\cite{simult}. 

First we need to review the physics behind confinement in QED$_3$~\cite{ConfQED3}. The potential between two static charges, as the separation between them, $r\to\infty$, behaves like
\be
V(r)=\frac{e^2}{8\pi} G(0) \ln(e^2r)+const+{\cal O}\left(\frac{1}{r}\right)\;.
\ee
This means that in the quenched theory, $G=1$, the potential is logarithmically confining. This is precisely the scenario we are considering. Bear in mind, however, that the situation changes drastically if loops of massless fermions are taken into account. Considering $N_f$ fermion families, we can easily prove that 
\be
G(k) = \frac{8\pi k}{8\pi k + e^2 N_f} \to 0 \qquad \mbox{as} \qquad k\to 0\;,
\ee
which is understandable on physical grounds, since massless fermions cost no energy to produce, and thus, they infinitely screen the static charges, sweeping away confinement. However, if the fermions within the loops acquire masses either dynamically or explicitly, they provide an effective screening, rendering $G(k)\ne 0$ and finite as $k\to 0$, and thus confinement is reinstated.

Confinement can be studied from the fermion propagator in the following form~\cite{ConfQED3}: Let us start defining the following space-averaged Schwinger function
\be
\Delta(t) = \int d^2x \int \frac{d^3p}{(2\pi)^3}e^{i(p_0x_0-\vec p\cdot \vec x)} \sigma_s(p)\;,\label{scw}
\ee
where
\be
\sigma_s(p)=\frac{F(p) M(p)}{p^2+M^2(p)}
\ee
is the scalar part of the fermion propagator. The Axiom of Reflection Positivity~\cite{ARP} establishes that if the propagator describes a stable particle,
\be
\Delta(t)\sim e^{-mt}
\ee
and $m$ corresponds to the mass of the particle. Such a behavior is found, for example, from the free fermion propagator. Actually, this test is performed in lattice simulations to obtain the mass of hadronic resonances.  On the other hand, when the propagator possesses complex singularities, $\Delta(t)$ is no longer positive definite, but behaves like
\be
\Delta(t)\sim e^{-m_1t}\cos(m_2 t+\delta)\;,
\ee
where $m_1$ and $m_2$ are the position of the said complex poles. In this case, the propagator describes a confined excitation. There are alternative formulations of this test~\cite{simult}.

\begin{center}
\begin{figure}[t]
\includegraphics[width=0.8\textwidth,angle=-90]{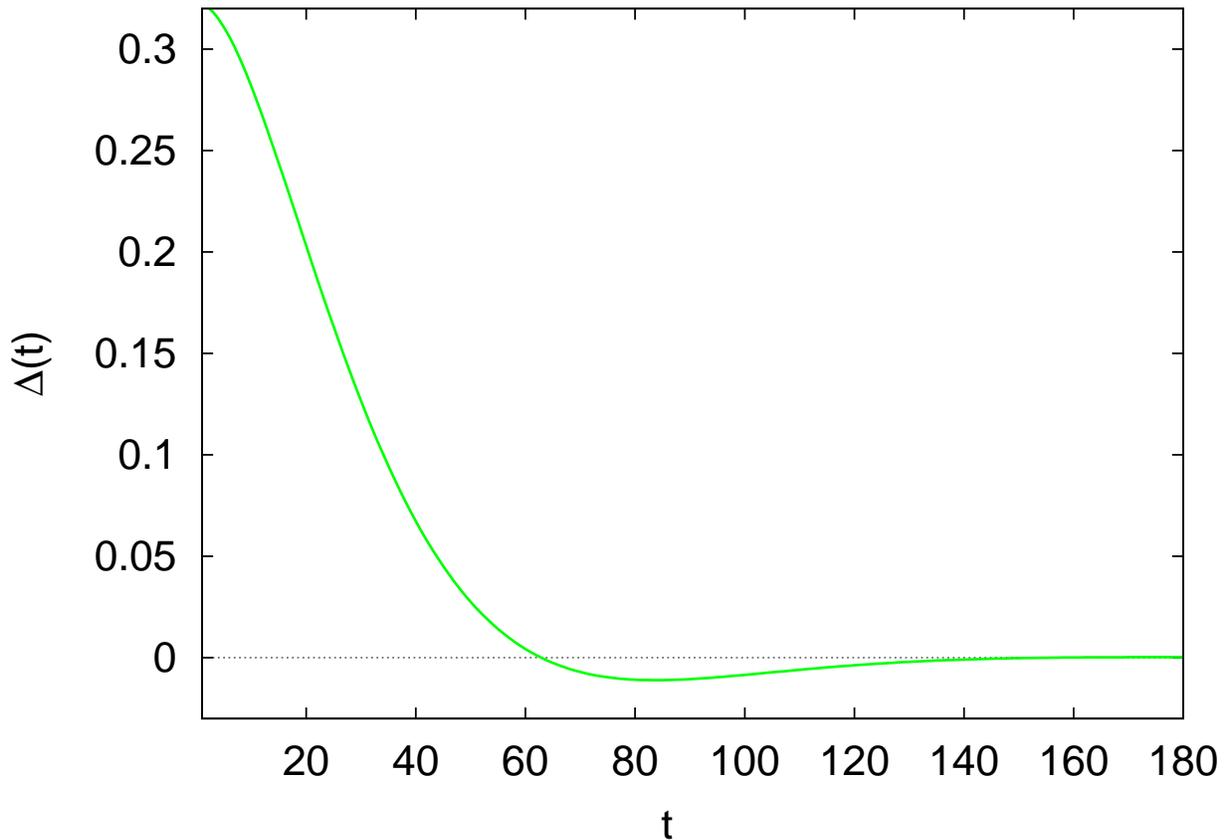}
\caption{Confinement test funtion $\Delta(t)$}
\label{fig:delta}
\end{figure}
\end{center}

\begin{center}
\begin{figure}[t]
\includegraphics[width=0.8\textwidth,angle=-90]{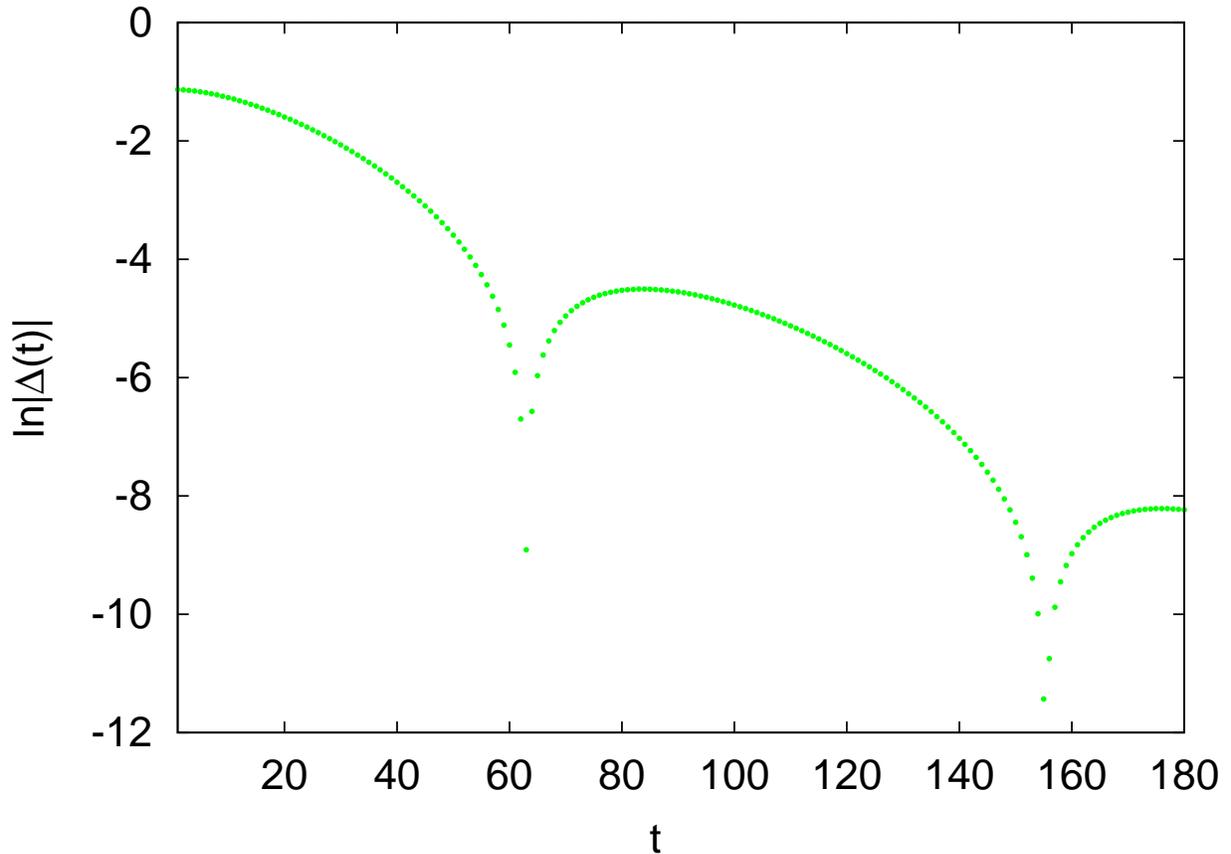}
\caption{Logarithm of the confinement test funtion $\Delta(t)$}
\label{fig:logdelta}
\end{figure}
\end{center}

If we compute the spatially averaged Schwinger function, eq.~(\ref{scw}), with the analytical form of our problem, eq.~(\ref{mp}), we obtain the curve shown in Fig.~(\ref{fig:delta}). We can see that, indeed, our solution induces an oscillatory behavior on the Schwinger function, which is more evident in Fig.~(\ref{fig:logdelta}), where we plot the logarithm of the absolute value of $\Delta(t)$. Each peak corresponds to a solution of $\Delta(t)=0$. Thus, we conclude that both DCSB and confinement are  present in our study.

A word of caution is at hand. The positiveness of the Green's function is a sufficient condition for confinement, but not necessary. The counterexample is QCD in (1+1) dimensions~\cite{QCD2}. There, the propagator exhibits complex singularities, but the color singlet meson bound state amplitudes vanish precisely at the fermion mass shell momenta. However, confinement is realized in such a theory via a failure of the so called cluster decomposition property~\cite{CDP}.

\section{Final Remarks}

Through these lectures, we have reviewed the dynamical generation of light quark mass. Such a mass is a measure of the confining energy of these quarks inside the nucleons, and is determined by the strong interactions of QCD. To understand the relationship between confinement and DCSB, we have worked out an neat example of a toy model of QCD which exhibits both these phenomena, QED$_3$. The SDE is the platform we have selected for this purpose, performing a truncation the tower of SDEs under simplifying assumptions. We have presented  a detailed  analytical solution to the gap equation of QED$_3$, and also, we have developed a numerical procedure to look for the dynamically generated fermion propagator. A confinement test performed on the propagator revealed that this chirally asymmetric solution to the SDE also describes a confined excitation.  This exercise points out the key features of strong interactions which reveal the origin of mass of the visible Universe.


\begin{theacknowledgments}
The author acknowledges the organizers of the XIII-MSPF, Maria Elena Tejeda-Yeomans, Carlos Calc\'aneo and the studes of USON, for their kind hospitality as well as valuable discussions with A. Ayala, A. Bashir and A. S\'anchez. He also acknowledges support from CONACyT, COECyT and CIC grants under projects 82230, and 4.22.
\end{theacknowledgments}

\end{document}